\documentstyle[prl,aps,preprint,tighten,floats,aps,amssymb]{revtex}
\input epsf.tex

\newcommand{\bmat}{\left(\begin{array}}
\newcommand{\emat}{\end{array}\right)}

\def\yzero{\smash{\hbox{$y\kern-4pt\raise1pt\hbox{${}^\circ$}$}}}

\def\-{\hphantom{-}}

\def\s2{\frac{1}{\sqrt2}}

\def\beq{\begin{equation}}
\def\eeq{\end{equation}}
\def\beqa{\begin{eqnarray}}
\def\eeqa{\end{eqnarray}}

\def\IF{\relax{\rm I\kern-.18em F}}
\def\II{\relax{\rm I\kern-.18em I}}
\def\IP{\relax{\rm I\kern-.18em P}}

\def\Dsl{\,\raise.15ex\hbox{/}\mkern-13.5mu D} 

\def\IC{\bf C}
\def\IZ{\bf Z}

\def\z2z2{$\IC^3/(\IZ_2\times\IZ_2)$}

\def\s{\sigma}

\def\z{\zeta}

\def\bo{{\raise-.3ex\hbox{\large$\Box$}}}               
\def\face{{\raise.2ex\hbox{$\displaystyle \bigodot$}\mskip-2.2mu \llap {$\ddot
        \smile$}}}                                      

\def\leftrightarrowfill{$\mathsurround=0pt \mathord\leftarrow \mkern-6mu
        \cleaders\hbox{$\mkern-2mu \mathord- \mkern-2mu$}\hfill
        \mkern-6mu \mathord\rightarrow$}       
\def\dvec#1{\vbox{\ialign{##\crcr
        \leftrightarrowfill\crcr\noalign{\kern-1pt\nointerlineskip}
        $\hfil\displaystyle{#1}\hfil$\crcr}}}           

\def\beq{\begin{equation}}
\def\eeq{\end{equation}}

\def\beqx{\begin{displaymath}}
\def\eeqx{\end{displaymath}}

\def\beqa{\begin{eqnarray}}
\def\eeqa{\end{eqnarray}}

\begin{document}
\draft
\date{August 30, 2001}
\title{
\normalsize
\mbox{ }\hspace{\fill}
\begin{minipage}{4cm}
UPR--951--T 
\end{minipage}\\[8ex]
{\Large On Type~II Superstrings in Less Than Four Dimensions
\\[1ex]}}

\author{Jens Erler and Gary Shiu}
\address{Department of Physics and Astronomy, University of Pennsylvania, \\
Philadelphia, PA 19104-6396, USA}

\maketitle

\thispagestyle{empty}

\begin{abstract}
We study gauge theories which are associated with classical vacua of 
perturbative Type~II string theory that allows for a conformal field theory
description.  We show that even if we compactify  
seven spatial dimensions (allowing for one macroscopic dimension to arise 
non-perturbatively) the Standard Model cannot be obtained from
the perturbative sector of Type~II superstrings.
Therefore, the construction of the Standard Model (or extensions thereof)
from M theory must involve fields that are non-perturbative from 
the Type~II perspective.  We also address the case of eight compact dimensions.
\end{abstract}

\newpage

One of the outstanding questions in string theory is {\it how\/} does it
describe nature. At present, we are unable to make definitive statements about
the underlying dynamics that selects our string vacuum. Nevertheless, one can 
use phenomenological constraints as guidelines to explore realistic string 
models in various calculable regimes.  The primary purpose of such explorations
is not to find the model which fully describes our world, but to 
examine the possibilities within the string theoretical framework.
 
In this respect, a perturbative limit of M theory which was largely unexplored 
is Type~II string theory (for earlier work in this direction, see, {\it e.g.}, 
References~\cite{DKV,kawai,dolan}). This is due to the no-go theorem in 
Ref.~\cite{DKV}, where it was shown on general grounds that Type~II string 
theory does not contain the Standard Model.  In this short note, we revisit 
this no-go theorem by compactifying this theory down to three dimensions.  
We also address compactifications to two dimensions.\footnote{The conformal 
field theory description of the heterotic string compactified on manifolds with
exceptional holonomy to three and two dimensions was considered in 
Ref.~\cite{Shatashvili}; that of Type~II strings on non-compact manifolds with
$G_2$ holonomy was recently studied in Ref.~\cite{eguchi}.}

Of course, our spacetime appears to be manifestly four-dimensional (rather
than three or lower dimensional), but as the string coupling becomes strong, 
a new dimension can open up.  In this limit, one may hope to recover an 
(approximately) Lorentz invariant four-dimensional spacetime. In fact, this 
transition from three to four dimensions was used in the proposal of
Ref.~\cite{Witten} as a possible solution to the cosmological constant problem
\footnote{A possible realization of this idea as M theory compactified
on a $Spin(7)$ manifold which asymptotes to 
a $G_2$ manifold  times $S^1$ 
was recently discussed in Reference~\cite{Spin7}.}.
Another motivation for studying this setup is that its M theory lift should
correspond to compactifications on seven-dimensional spaces which may not even
have a geometrical interpretation\footnote{For geometrical compactifications 
on seven-manifolds with $G_2$ holonomy and their Type~II orientifold cousins,
see References~\cite{gtwo,GSS} and~\cite{CSU}, respectively. M-theory
models on Calabi-Yau threefolds times $S^1/Z_2$~\cite{horava} and
asymmetric orbifolds~\cite{asyorb} have also been considered~\cite{Dine}.}. 

We will show that there is enough room to accommodate the Standard Model
gauge group $SU(3) \times SU(2) \times U(1)$ and matter fields
transforming in appropriate representations under the
non-Abelian gauge groups. However, the $U(1)$ quantum numbers 
carried by the candidate quarks and leptons cannot be identified with
hypercharge. Therefore, the Standard Model, if obtained from M theory, 
must contain sectors that are non-perturbative from 
the Type~II point of view.

The argument is similar in spirit to that in Ref.~\cite{DKV}.
The only requirements are superconformal invariance and the decomposition
of the super-stress tensor $T$ into 
a sum of contributions from the Minkowski space-time
and the internal conformal field theory.
There is no need to assume any particular compactification
scheme, orbifold or otherwise, space-time supersymmetric or not.
The internal conformal field theory is not required to have any
geometrical interpretation.

Our starting point is a classical background of 
Type~II string theory which is defined
by a conformal field theory with local
$(1,1)$ world-sheet supersymmetry.
The matter (primary) superfields are defined through their operator product 
expansions (OPEs) with the (left-moving) holomorphic super-stress tensor
$T(z,\theta) = T_F (z) + \theta T_B (z)$ and likewise with its
anti-holomorphic counterpart
$\overline{T}(\overline{z},\overline{\theta})
=\overline{T}_F (\overline{z}) + \overline{\theta} ~\overline{T}_B (\overline{z})$ where
$(z_, \theta)$ and $(\overline{z},\overline{\theta})$
are the holomorphic and anti-holomorphic coordinates
parametrizing the two-dimensional superspace.
The OPE of $T(z,\theta)$ with itself is given by
\begin{equation}\label{OPE-TT}
T(z_1,\theta_1) \cdot T(z_2, \theta_2) =
\frac{\frac{1}{4}\hat{c}}{z_{12}^3}
+  \frac{3 \theta_{12}}{2 z_{12}^2} ~T (z_2,\theta_2)
+ \frac{\frac{1}{2}}{z_{12}} ~D_2 T 
+ \frac{\theta_{12}}{z_{12}} ~\partial_2 T 
+ \dots ~,
\end{equation}
where $z_{12} = z_1 - z_2 - \theta_1 \theta_2$ and
$\theta_{12} = \theta_1 - \theta_2$.
Since the ghosts for the local super-reparametrization invariance
contribute a central charge $\hat{c}_{ghost}=-10$,
the total superconformal anomaly can cancel only if
the central charge of the superconformal field theory (SCFT) describing
the matter superfields has $\hat{c}=10$.

Let us decompose the superconformal field theory into 
a direct product of a D-dimensional 
space-time and an internal superconformal field
theory. The super-stress tensor can be written as
$T(z,\theta) \equiv T^{D}(z,\theta) + T^{\rm int} (z,\theta)$
where $T^D$ and $T^{\rm int}$ anti-commute with each other.
The central charge of the space-time SCFT is $\hat{c}=D$, and hence
$\hat{c}_{\rm int}=10-D$.
For $D=4$, $\hat{c}_{\rm int}=6$, it was shown in Ref.~\cite{DKV} that the 
central charge of the internal conformal field
theory is not large enough to accommodate the Standard Model
gauge group, if there are massless fermions in the
appropriate representations of $SU(3) \times SU(2)$ playing the
role of quarks and leptons.  It was also shown there~\cite{DKV} that
chiral fermions transform only under gauge symmetries from one 
side of the superstring, so that one can focus, say, 
on the left-movers.

In three dimensions, the situation seems better.
The central charge of the internal conformal
field theory is $\hat{c}_{\rm int}=7$. As in References~\cite{DKV,kawai},
one can give a complete list of 3D gauge groups allowed in
perturbative Type~II string theory by considering all
possible semi-simple Lie algebras with central charge $\hat{c} \leq 7$.
Let $J^a$ be a holomorphic supercurrent of the Kac-Moody algebra.
The OPEs of $J (z,\theta)$ with $T (z,\theta)$ and itself
are given by,
\begin{eqnarray}
T(z_1,\theta_1) \cdot J^a (z_2,\theta_2)
&=& \frac{\theta_{12}}{z_{12}^2} J^a (z_2,\theta_2)
+\frac{\frac{1}{2}}{z_{12}} D_2 J^a (z_2,\theta_2)
+ \frac{\theta_{12}}{z_{12}} \partial_2 J^a (z_2,\theta_2) + \dots~,
\nonumber \\
J^a (z_1,\theta_1) \cdot J^b (z_2,\theta_2) &=&
\frac{1}{2 z_{12}} \cdot k \delta^{ab} {\bf 1}
+ \frac{\theta_{12}}{z_{12}} \cdot i f^{abc} J^c (z_2, \theta_2) + \dots~,
\end{eqnarray}
where $f^{abc}$ are the structure constants of the semi-simple Lie group.
Unitarity requires the Kac-Moody level, $k_i$, and the super-Kac-Moody level,
$\hat{k}_i = k_i - C_A$, of each group factor, $G_i$, to be non-negative 
integers. ($C_A$ denotes the eigenvalue of the quadratic Casimir operator
in the adjoint representation.)  
In terms of these and the dimension of $G_i$, the central charge is given 
by~\cite{Kac}:
\begin{equation}
\hat{c} (G_i) = \frac{d(G_i)}{3} + \frac{2 d(G_i)}{3} 
\frac{k_i - C_A}{k_i}~.
\end{equation}
Descendant (secondary) fields do not give rise to massless states.
Thus, Standard Model fields must be described by primary fields.
However, primary fields transforming under representations
of $G_i$ (other than the identity representation) can only be present
if $\hat{k}_i > 0$.  Applying this to the Standard Model we consider
the super-Kac-Moody algebra corresponding to
$SU(3)_{k_3} \times SU(2)_{k_2} \times U(1)_{k_1}$
\footnote{At the level of the superconformal algebra, 
the level of a $U(1)$ Kac-Moody 
algebra is not well
defined. However, in many cases, it can be read off from the anomaly
polynomial \cite{erler}.
Here, we simply mean the normalization of the $U(1)$ charge.}.
The central charges $\hat{c}_3$ and $\hat{c}_2$ of $SU(3)_{k_3}$ and
$SU(2)_{k_2}$ are given by
\begin{eqnarray}
\hat{c}_3 &=& \frac{8}{3} + \frac{16}{3}~\frac{k_3 - 3}{k_3} ~, \\
\hat{c}_2 &=& 1 + 2~\frac{k_2 - 2}{k_2} ~,
\end{eqnarray}
while $U(1)_{k_1}$ contributes central charge $\hat{c}_1=1$ for any $k_1$.
It was the observation that~\cite{DKV}
\begin{equation}
\hat{c}_3 + \hat{c}_2 + \hat{c}_1 \geq 20/3,
\end{equation}
while $\hat{c}^{\rm int} = 6$ that excluded the perturbative Type~II 
superstring from further phenomenological considerations.
One possible way out, as noted in~\cite{DKV}, is to give up 
one spatial dimension.
We will now show that the situation is actually significantly worse than 
that.

Compactification down to three dimensions implies $\hat{c}^{\rm int} = 7$,
and therefore
\begin{equation}
\frac{16}{3} \frac{k_3-3}{k_3} + 2~\frac{k_2 -2}{k_2}  \leq \frac{7}{3}~.
\end{equation}
The only allowed values are $(k_2,k_3)=(3,4)$ or $(4,4)$.
While the latter case saturates the bound, in the former there must be 
an additional superconformal field theory with $\hat{c} = 1/3$.
However, the central charge for unitary superconformal field
theories with $\hat{c}<1$ is highly restricted. The only
allowed values are~\cite{Kac},
\begin{equation}
\hat{c} = 1 - \frac{8}{m(m+2)} ~, \quad \quad \quad m=2,3,4,\dots,
\end{equation}
and combinations thereof.  Since $\hat{c} = 1/3$ is not allowed,
the case $(k_2,k_3)=(3,4)$ must be rejected. We are left with
$(k_2,k_3)=(4,4)$, with no room for an extra conformal field theory.
Note that the gauge couplings for the higher level models 
are given in terms of the string coupling
\begin{equation}
g_{i}^2 k_i = g_s^2~.
\end{equation}
Since $SU(3)$ and $SU(2)$ are realized at the same Kac-Moody level,
they have the same gauge coupling at the string scale --- indicating
the unification of the $SU(3)$ and $SU(2)$ gauge couplings slightly
below the string scale consistent with observation. This is in contrast 
with the perturbative heterotic string case where the Kac-Moody levels
are generally unconstrained.

The conformal dimension of a primary field corresponding to the highest 
weight representation $r$ is
\begin{equation}
h_r = \frac{C_r}{2k+C_A}~,
\end{equation}
where $C_r$ is the quadratic Casimir of $r$. Hence, for the fundamental 
representation ${\bf N}$ of $SU(N)_k$,
\begin{equation}
h_{\bf N} = \frac{N^2-1}{2N(k+N)}~.
\end{equation}
The conformal dimension of a primary field carrying charge $Q$ with respect
to an Abelian Kac-Moody factor realized at level $k_1$ is given by
\begin{equation}
h_{Q} = \frac{1}{2k_1}~ Q^2~,
\end{equation}
where $k_1$ serves as a normalization constant.

Identifying the $U(1)$ with hypercharge the conformal weights of various
Standard Model fields are:
\begin{tabbing}
\qquad \qquad \qquad\= \underline{Field} \qquad \qquad\=
$\underline{SU(3) \times SU(2) \times U(1)_Y}$ \qquad  \qquad\=
\underline{Conformal weight $h$} \\[0.8ex]
\> $Q_L$ \> \quad \qquad \quad $({\bf 3},{\bf 2})_{\frac{1}{6}}$  \> \qquad
$\frac{53}{168} + \frac{1}{72 k_1}$ \\[0.8ex]
\> $U$ \> \quad \qquad \quad $(\overline{\bf 3},{\bf 1})_{\frac{1}{3}}$  \> \qquad
$\frac{4}{21} + \frac{1}{18 k_1}$ \\[0.8ex]
\> $D$ \> \quad \qquad \quad $(\overline{\bf 3},{\bf 1})_{-\frac{2}{3}}$  \> \qquad
$\frac{4}{21} + \frac{2}{9 k_1}$ \\[0.8ex]
\> $L$ \> \quad \qquad \quad $({\bf 1},{\bf 2})_{-\frac{1}{2}}$  \> \qquad
$\frac{1}{8} + \frac{1}{8 k_1}$ \\[0.8ex]
\> $H_U,H_D$ \> \quad \qquad \quad $({\bf 1},{\bf 2})_{\pm \frac{1}{2}}$  \> \qquad
$\frac{1}{8} + \frac{1}{8 k_1}$ \\[0.8ex]
\> $E$ \> \quad \qquad \quad $({\bf 1},{\bf 1})_{1}$  \> \qquad
$\frac{1}{2 k_1}$ \\[0.8ex]
\end{tabbing}
The conformal weight of a massless field is $h=\frac{1}{2}$.
It is easy to see that there is no choice of the normalization factor
$k_1$ such that all Standard Model fields are massless. 

The reason that a $\hat{c}_{int} = 7$ conformal field theory fails 
to describe the Standard Model is that it {\it barely\/} contains 
its gauge groups plus quarks and weak doublets.  Therefore, particles 
with the same non-Abelian gauge quantum numbers must have the same
hypercharge (up to a sign) to be simultaneously massless.
This implies that the up and down quark singlets are predicted
to carry the same amount of $U(1)$ charge and hence this $U(1)$ gauge field
cannot be hypercharge.  

It is interesting to note that if we 
compactify Type~II string theory
down to {\it two} dimensions, we have 
$\hat{c}_{int}=8$ which
can accommodate an extra $U(1)^{\prime}$.
The complete Standard Model spectrum can be massless as long
as it is supplemented by the appropriate $U(1)^{\prime}$ charges.
It turns out, however, that these charges are incompatible with the
Yukawa structure of the Standard Model.
One can categorize the possible gauge groups for
$\hat{c}_{int}=8$:
\begin{itemize}
\item (i) $SU(3)_4 \times SU(2)_4 \times U(1) \times U(1)^{\prime}$
\item (ii) $SU(3)_4 \times SU(2)_4 \times U(1) \times 
\left( \hat{c}=1~ {\mbox{SCFT}} \right)$
\item (iii) $SU(3)_4 \times SU(2)_3 \times U(1) \times \left( \hat{c}=\frac{2}{3} ~
{\mbox{SCFT}} \right) \times \left(\hat{c}=\frac{2}{3} ~
{\mbox{SCFT}} \right)$ 
\item
(iv) $SU(3)_4 \times SU(2)_6 \times U(1) \times \left( \hat{c}=\frac{2}{3} 
~{\mbox{SCFT}} \right)$
\item (v) $SU(3)_5 \times SU(2)_5 \times U(1)$ 
\item (vi) $SU(3)_6 \times SU(2)_3 \times U(1)$
\end{itemize}
It is easy to see that (v) and (vi) are ruled out by the same reasoning
that the $SU(3)_4 \times SU(2)_4 \times U(1)$ case was ruled out for 
$\hat{c}_{int}=7$. The conformal weights of primary fields in
$\hat{c}<1$ SCFT are highly restricted. One can show that 
for (iii) and (iv), there is no
normalization constant $k_1$ such that all Standard Model fields are
massless. 
A large class of 
$\hat{c}=1$ SCFTs have been studied in Reference~\cite{c=1}, but
it is not known whether this represents the complete classification.
We do not consider case (ii) in detail; however, it seems very unlikely
that the Standard Model can be constructed.

Let us note that it is even more difficult
to embed a Grand Unified Theory (GUT) or a model with 
partial unfication
({\it e.g.}, the Pati-Salam model) within perturbative
Type~II string theory. The  
GUT gauge group with the smallest dimension $d(G)$ is $SU(5)$ which
has $d(G)=24$, hence $\hat{c}_{int} \geq 8$.
In the case of the Pati-Salam model, the central charge of the
$SU(4)$ factor is at least $7$ if we require the 
fundamental representation of $SU(4)$ to be unitary (so that 
there are quarks upon breaking the Pati-Salam gauge group 
down to the Standard Model).
There is not enough room even to accommodate the weak $SU(2)$.
Similarly, the left-right model cannot be obtained as well.

As we go from three to four dimensions, we are in
the strong coupling regime
so that strictly speaking, we cannot trust a perturbative Type~II
calculations. However, with supersymmetry, it is likely that the
perturbative massless spectrum would stay massless even at strong
coupling. Certainly, there are additional massless states, 
including the D0-branes whose masses go as $1/R_{11}$ where $R_{11}$
is the size of the eleventh dimensions.

A possible way of evading the no-go theorem presented here is that
it may not always be possible to decompose 
the super-stress tensor $T$ into
a sum of $T^D$ and $T^{\mbox{\footnotesize int}}$ 
which commute with each other.
For example, this may be the case in a warped compactification where
the $D$-dimensional spacetime metric depends on the coordinates 
in the $(10-D)$ compactified dimensions. It would be interesting
to see if this may help in relaxing the constraints in
embedding the Standard Model in perturbative Type~II string theory.

So far we restricted our attention to perturbative spectra of Type~II string 
theory described by a CFT.  However, these spectra may be supplemented by 
non-perturbative (solitonic) states even at arbitrarily weak string 
coupling, when the compactification manifold is singular~\cite{duality}. 
These correspond to points in the moduli space where the associated 
conformal field theories are badly behaved~\cite{aspinwall}.
On the Type~IIA (IIB) side, the non-perturbative states 
are obtained from even (odd) D-branes wrapped 
around even (odd) collapsing cycles\footnote{It is not clear, however,
how this way of obtaining non-Abelian gauge symmetries from singularities
can be understood in a non-geometrical setup, such as asymmetric orbifolds.}. 
In this case, the gauge coupling is 
independent of the string coupling, but depends on the geometric moduli. 
It is interesting to note, that 
if the $U(1)_Y$ is a linear combination of a perturbative and
a non-perturbative $U(1)$ gauge field, then
our argument based on the conformal weights does not apply.
More generally, (parts of) the Standard Model gauge group can be a diagonal 
subgroup of a perturbative and a non-perturbative gauge group. This may lead 
to a field theoretical realization \cite{matt} of the
generation-changing phase transitions~\cite{KS,Park,CSU}.

\acknowledgments

We thank Mirjam Cveti\v c, Asad Naqvi, Joseph Polchinski, Koenraad Schalm,
Matthew Strassler, and Henry Tye 
for discussions and comments.  This work was supported
by the DOE grants DE-FG02-95ER40893 and DE-EY-76-02-3071, as well as the 
University of Pennsylvania School of Arts and Sciences Dean's fund.




\newpage

\begin{thebibliography}{99}

\bibitem{DKV}
L.~J.~Dixon, V.~Kaplunovsky, and C.~Vafa,
Nucl.\ Phys.\ B {\bf 294}, 43 (1987).

\bibitem{kawai}
H.~Kawai, D.~C.~Lewellen, and S.~H.~Tye,
Phys.\ Lett.\ B {\bf 191}, 63 (1987).

\bibitem{dolan}
R.~Bluhm, L.~Dolan, and P.~Goddard,
Nucl.\ Phys.\ B {\bf 289}, 364 (1987); \\
W.~Lerche, B.~E.~Nilsson, and A.~N.~Schellekens,
Nucl.\ Phys.\ B {\bf 294}, 136 (1987).

\bibitem{Shatashvili}
S.~L.~Shatashvili and C.~Vafa,
hep-th/9407025.

\bibitem{eguchi}
T.~Eguchi and Y.~Sugawara, hep-th/0108091.

\bibitem{Witten}
E.~Witten,
Mod.\ Phys.\ Lett.\ A {\bf 10}, 2153 (1995).

\bibitem{Spin7}
M.~Cvetic, G.~W.~Gibbons, H.~Lu and C.~N.~Pope,
hep-th/0103155.

\bibitem{gtwo} 
B.~S.~Acharya,
{hep-th/0011089};
{hep-th/0101206}; \\
M.~Atiyah, J.~Maldacena, and C.~Vafa,
{hep-th/0011256}; \\
M. Atiyah and E. Witten, hep-th/0107177; \\
M.~Cveti$\check{\rm c}$, G.~W.~Gibbons, H.~L\" u, and C.~N.~Pope,
{hep-th/0106177};\\
E.~Witten,
hep-th/0108165.

\bibitem{GSS}
B.~R.~Greene, K.~Schalm and G.~Shiu,
hep-th/0010207,
J.\ Math.\ Phys. {\bf 42}, 3171 (2001).


\bibitem{CSU}
M.~Cveti\v c, G.~Shiu, and A.~M.~Uranga,
hep-th/0107143; hep-th/0107166.

\bibitem{horava}
P.~Horava and E.~Witten,
Nucl.\ Phys.\ B {\bf 460}, 506 (1996) and {\bf 475}, 94 (1996); \\
B.~A.~Ovrut,
hep-th/9905115 and references therein.


\bibitem{asyorb}
K.~S.~Narain, M.~H.~Sarmadi, and C.~Vafa,
Nucl.\ Phys.\ B {\bf 288}, 551 (1987).

\bibitem{Dine}
M.~Dine and E.~Silverstein,
hep-th/9712166.



\bibitem{Kac}
V.~G.~Kac and I.~T.~Todorov,
Commun.\ Math.\ Phys.\  {\bf 102} (1985) 337.

\bibitem{erler}
J.~Erler,
Nucl.\ Phys.\ B {\bf 475}, 597 (1996).


\bibitem{c=1}
L.~J.~Dixon, P.~Ginsparg, and J.~A.~Harvey,
Nucl.\ Phys.\ B {\bf 306}, 470 (1988).

\bibitem{duality}
E.~Witten,
Nucl.\ Phys.\ B {\bf 443}, 85 (1995); \\
C.~M.~Hull and P.~K.~Townsend,
Nucl.\ Phys.\ B {\bf 451}, 525 (1995).

\bibitem{aspinwall}
P.~S.~Aspinwall,
Phys.\ Lett.\ B {\bf 357}, 329 (1995); Phys.\ Lett.\ B {\bf 371}, 231 (1996);
hep-th/9611137 and references therein.

\bibitem{matt} M.J.~Strassler, in preparation.


\bibitem{KS}
S.~Kachru and E.~Silverstein,
Nucl.\ Phys.\ B {\bf 504}, 272 (1997).

\bibitem{Park}
B.~A.~Ovrut, T.~Pantev and J.~Park,
JHEP {\bf 0005}, 045 (2000).


\end{thebibliography}
\end{document}